\documentstyle[12pt,epsf,epsfig]{article}
\def\be{\begin{equation}}
\def\ee{\end{equation}}
\def\ba{\begin{eqnarray}}
\def\ea{\end{eqnarray}}

\def\bdm{\begin{displaymath}}
\def\edm{\end{displaymath}}
\def\la{~\mbox{\raisebox{-.6ex}{$\stackrel{<}{\sim}$}}~}
\def\ga{~\mbox{\raisebox{-.6ex}{$\stackrel{>}{\sim}$}}~}
\def\bq{\begin{quote}}
\def\eq{\end{quote}}

 at 10truept

\newcommand{\beq}{\begin{equation}}
\newcommand{\eeq}{\end{equation}}
\newcommand{\beqa}{\begin{eqnarray}}
\newcommand{\eeqa}{\end{eqnarray}}

\def\la{~\mbox{\raisebox{-.6ex}{$\stackrel{<}{\sim}$}}~}
\def\ga{~\mbox{\raisebox{-.6ex}{$\stackrel{>}{\sim}$}}~}

\def\ltap{\ \raise.3ex\hbox{$<$\kern-.75em\lower1ex\hbox{$\sim$}}\ }
\def\gtap{\ \raise.3ex\hbox{$>$\kern-.75em\lower1ex\hbox{$\sim$}}\ }
\def\gl{\ \raise.5ex\hbox{$>$}\kern-.8em\lower.5ex\hbox{$<$}\ }
\def\roughly#1{\raise.3ex\hbox{$#1$\kern-.75em\lower1ex\hbox{$\sim$}}}

\begin{document}

\thispagestyle{empty}
\begin{flushright}
arXiv:0706.1977 [astro-ph]\\
June 2007
\end{flushright}
\vspace*{1cm}
\begin{center}
{\Large \bf Challenging the Cosmological Constant}\\

\vspace*{1.5cm} {\large Nemanja Kaloper\footnote{\tt
kaloper@physics.ucdavis.edu}}\\
\vspace{.5cm} {\em Department of Physics, University of
California, Davis,
CA 95616}\\
\vspace{.05cm} \vspace{1.3cm} ABSTRACT
\end{center}
We outline a dynamical dark energy scenario whose signatures may
be simultaneously tested by astronomical observations and
laboratory experiments. The dark energy is a field with slightly
sub-gravitational couplings to matter, a logarithmic
self-interaction potential with a scale tuned to $\sim 10^{-3} \,
{\rm eV}$, as is usual in quintessence models, and an effective
mass $m_\phi$ influenced by the environmental energy density. Its
forces may be suppressed just below the current bounds by the
chameleon-like mimicry, whereby only outer layers of mass
distributions, of thickness $1/m_\phi$, give off appreciable long
range forces. After inflation and reheating, the field is
relativistic, and attains a Planckian expectation value before
Hubble friction freezes it. This can make gravity in space
slightly stronger than on Earth. During the matter era, interactions
with nonrelativistic matter dig a minimum close to the Planck scale.
However, due to its sub-gravitational matter couplings the field
will linger away from this minimum until the matter energy density
dips below $\sim 10^{-12} \, {\rm eV}^4$. Then it starts to roll
to the minimum, driving a period of cosmic acceleration.
Among the signatures of this scenario may be dark energy equation
of state $w \ne -1$, stronger gravity in dilute mediums, that may
influence BBN and appear as an excess of dark matter, and
sub-millimeter corrections to Newton's law, close to the present
laboratory limits.

\vfill \setcounter{page}{0} \setcounter{footnote}{0}
\newpage

Understanding cosmic acceleration is the deepest problem of modern
cosmology. It has profound implications both for fundamental
physics and for the fate of the universe \cite{wein}. A range of
ideas have been pursued to explain the acceleration, and to date
in all of them, one is forced to fine tune some dimensional scales
to accommodate cosmic acceleration {\it now}. This yields the `Why
Now' problem, which may be taken as a clue that we are missing
something important in the formulation of the problem
\cite{whynow}. To compound the puzzle, to date we have noted other
curious coincidences, such as the near matches between the scale
of the cosmological constant, the dark matter density, the neutrino
mass, and the laboratory limits on gravitational force, which are
all controlled by a length scale of about a millimeter. While
these may simply be numerical accidents, it is interesting to
probe for deeper connections between them. We can pursue this by
formulating models where cosmic acceleration has other direct
observable consequences, as exemplified in \cite{beane}-\cite{lawrence}.

The main problem in building such models is the range of mass
scales which one needs for nontrivial dynamics. For example, to
have a dynamical dark energy instead of the cosmological constant one
needs ultralight degrees of freedom, say scalars, with masses
$m_\phi \la H_0 \sim 10^{-33} {\rm eV}$. These must couple to
matter {\it significantly} more weakly than gravity to avoid
conflicts with Solar System tests \cite{quint}. On the other hand,
laboratory tests constrain new fields to be heavier than about
$10^{-3} {\rm eV}$, if they couple to matter gravitationally
\cite{laboratory}. So to make dark energy detectable in laboratory
searches and consistent with long range gravity, we need models
where its mass {\it changes} by at least {\it thirty} orders of
magnitude between the Earth and the extragalactic space. Indeed,
if the masses of dark fields are fixed by the current laboratory
bounds, we could integrate them out at scales below their masses
and end up with dark energy practically indistinguishable from the
pure cosmological constant, without a direct link to laboratory
phenomena.

In this note we will outline a model of quintessence which may be
within reach of future terrestrial searches for sub-millimeter
corrections to Newton's law of gravity. It controls cosmology at
largest scales with a very weak potential, logarithmic in the
field value. Yet at shorter scales, due to large environmental
masses as in \cite{dano,dapol,khuwe}, this field could decouple at
the scales probed by current laboratory tests, but perhaps just barely, so that it
could be revealed by future probes. Its signatures, in addition to
possible sub-millimeter gravitational effects, would include an equation
of state $w\ne -1$, distinguishing it from the cosmological
constant, stronger gravity in less dense mediums, which can
influence BBN, and induce a weak spatio-temporal variation of
Newton's constant, affecting structure formation and possibly
simulating an excess of dark matter abundance over its actual
density. This model could therefore be a useful benchmark for
future observational explorations of the signatures of dark
energy.

We start our discussion with a review of the mechanisms that make
the masses of fields dependent on the medium in which they
propagate \cite{dano}-\cite{gukhu}. They may provide a way around
the usual decoupling argument, and are most simply formulated for
models where the scalar couples to matter
universally\footnote{Wider classes of models where the coupling
changes from species to species were studied in \cite{dapol}.}, by
interaction Lagrangians ${\cal L}_{matter}(g^{\mu\nu} e^{-2 \alpha
\phi/M_4}, \Psi)$ like a Brans-Dicke field. In these cases, the
effective potential controlling the propagation of a field in a
medium is given by
\be
V_{eff}(\phi) = V(\phi) - T^\mu{}_\mu \, e^{\alpha_w \phi/M_4} \, ,
\label{effpot}
\ee
where $V(\phi)$ is the potential in the vacuum and $T^\mu{}_\mu
\propto - \rho$ is the trace of the stress energy of the
environment\footnote{Our conventions are $M_4^2 G_{\mu\nu} =
T_{\mu\nu}$ for the Einstein's equations and $\delta S_{matter} =
\frac12 \int d^4x \sqrt{\bar g_4} \, \bar T^{\mu\nu} \,\delta \bar
g_{\mu\nu}$ for the stress energy tensor in the Brans-Dicke frame.
We will define Einstein frame components of $\bar T_{\mu\nu}$ by
$\rho, p = e^{3(1+w) \alpha \phi/M_4} \bar \rho, \bar p$,
respectively, for reasons of simplicity, to be noted shortly.}.
The Planck mass $M_4$ and the dimensionless quantity $\alpha_w =
(1-3w) \alpha$ parameterize the couplings of the scalar to matter.
Thus in stationary matter distributions, the minimum of the field
$\phi$ is at $\phi_*$, where $\partial_\phi V_{eff}(\phi_*) = 0$.
The effective mass governing the dynamics of the field
fluctuations about  this environmental minimum $\phi_*$ is
$m^2_\phi = \partial_\phi{}^2 \, V_{eff}(\phi_*)$. In
distributions of matter with energy density $\rho$ and pressure
$p$, $T^\mu{}_\mu = - (\rho - 3 p)$, setting how $\phi_*$ and
$m^2_\phi$ will depend on the energy density of the environment.
As the energy density changes, so will the location of the minimum
$\phi_*$. Over cosmological time scales, the evolution of the zero
mode is governed by
\ba &&3M_4^2 H^2 = \frac{\dot \phi^2}{2} + V + \rho \, e^{\alpha_w
\phi/M_4} \, ,
\label{hubble} \\
&&\dot \rho + 3(1+w) H\rho  = 0 \label{gas} \, , \\
&&\ddot \phi + 3H \dot \phi + \frac{\partial V_{eff}}{\partial
\phi} = 0 \label{fieldeq} \, , \ea
which come from the Einstein's equations and the $\phi$ field
equation in homogeneous and isotropic, spatially flat FRW
universes, that are a good approximation for our universe from just
after the beginning of inflation onwards. The simplicity of the source
terms is ensured by our conventions. Clearly, $\phi_*$ is not an
exact solution to these equations, but will be a good
approximation over time scales $t \ll 1/H$, if $m_\phi > H$. From these equations,
we can immediately find the condition when $\phi$ can yield cosmic
acceleration. Acceleration is {\it not} automatic: even if $\rho$
is propping $\phi$ up on a slope of $V$, it changes due to
cosmic expansion, and the field $\phi$ may slide down $V$ too
fast to support cosmic acceleration over a Hubble time. Indeed, we
can check immediately that for the example of nonrelativistic matter, 
if $\phi$ sits in the minimum
of $V_{eff}$ the total energy density changes according to $ \dot
H =  - \frac{\dot \phi^2}{2M_4^2} - \frac{\rho}{2M_4^2} e^{\alpha
\phi/M_4} \simeq - \frac32 H^2$, which is clearly too fast to
support acceleration.

The criteria for acceleration can be formulated by generalizing
inflationary slow roll parameters to arbitrary fluids. Using
critical energy density $\rho_{cr} = 3M^2_4 H^2$ we see that the
universe will accelerate if
\be \epsilon = |\frac{\dot \rho_{cr}}{H\rho_{cr}}| < 1 \, .
\label{epcrit} \ee
Acceleration will last an e-fold or more if
\be
\eta = |\frac{\dot \epsilon}{3H\epsilon} | < 1 \, ,
\label{etacrit}
\ee
sustaining potential dominance for at least a Hubble time. We can
now find the conditions for acceleration as follows. Suppose first
that $m_\phi > H$. Then Eq. (\ref{fieldeq}) tells that $\phi$ will
rapidly settle into the environmental minimum $\phi_*$, during a
time scale $1/m_\phi$ over which the Hubble friction is
negligible. The direct evaluation of the $\epsilon$ parameter then
shows that $\epsilon = [\dot \phi^2 + (1+w) \rho e^{\alpha_w
\phi/M_4}]/{V}$. Then approximating $\phi \sim \phi_*$,
differentiating $\frac{\partial V_{eff}}{\partial \phi_*}$ with
respect to time and squaring it yields $\dot \phi^2 \simeq
9(1-3w)^2 \alpha^2_w \frac{H^2 \rho e^{\alpha_w \phi/M_4}}{M_4^2
m_\phi^4} \rho e^{\alpha_w \phi/M_4}$. Further using $\epsilon \la
1$ and $m_\phi > H$ yields $\dot \phi^2 < 9(1-3w)^2 \alpha^2_w
\rho e^{\alpha_w \phi/M_4}$. Thus generically we can neglect $\sim
\dot \phi^2$ terms in $\epsilon$, yielding $\epsilon \simeq (1+w)
\rho e^{\alpha \phi/M_4}/{V}$. Using this to evaluate $\eta$ in
the limit $m_\phi > H$, we find that $\eta \simeq  (1+w)^2$. So
when $m_\phi > H$, cosmic acceleration won't last longer than only a
fraction of an e-fold unless the environment obeys $|w+1| < 1$.
But that means that an agent other than $\phi$ plays the role of
dark energy, and $\phi$ is merely a spectator. Hence if $\phi$ is
to be dark energy at any time, we must have
\be
m_\phi \la H \, ,
\label{mdark}
\ee
over the relevant scales. In particular, for our $\phi$ to explain
cosmic acceleration now we need $m_\phi \la H_0$ at horizon
scales, unless we introduce some other dark energy by hand.

Now, the vacuum potential $V$ must satisfy some conditions in
order to allow for a dynamical setup which won't violate
experimental bounds on deviations from General Relativity, while
still yielding something non-trivial. The allure of the chameleon
mechanism is the environmental screening of the long range forces
from matter interior to the mass distributions \cite{khuwe}.
Namely, inside masses the environmental mass of the field $m_\phi$
is shifted up to a value much larger than in the vacuum, and so
the chameleon forces of particles inside the distributions acquire
efficient Yukawa suppressions, by the exponent of the depth
of the source particle inside the mass distribution, in the units of $1/m_\phi$.
The suppressions die out for
particles in the outer layer of the mass, of thickness roughly
$\sim 1/m_\phi(\phi)$ (which may have to be evaluated at some interpolating value
of $\phi$ nearer to the boundary of the matter distribution,
rather that its value $\phi_*$ in the core, to account 
for the variation of the homogeneous field
mode through the matter distribution). 
This yields the net scalar force suppression
relative to gravity by a factor of roughly $\sim m_\phi^{-1}/{\cal
R}$, where ${\cal R}$ is the size of the source, even if the
scalar is ultralight outside of the masses \cite{khuwe}. Clearly,
the smaller the source, the less the suppression, and this is why
for laboratory experiments, which work at a millimeter scale,
this still translates to roughly $m_\phi \ga 10^{-3} \, {\rm eV}$, for couplings of the order of
${\cal O}(1) \times M^{-1}_4$. Another important phenomenon
concerns the effective gravitational coupling of matter. One sees
immediately that environmental minima must obey $| \alpha \Delta
\phi_*| < M_4$ over a wide range of scales, where $\Delta \phi_*$
is the shift of $\phi_*$ with the change of $\rho$. Otherwise, the
effective gravitational coupling $G_{N \, eff} \sim
\frac{1}{M_4^2} \exp(\alpha_w \phi_*/M_4)$ would change too much
between the laboratory and, say, the atmosphere \cite{khuwe}. We should mention that
the bounds from astrophysical gravitational fields may be weaker because
of various model-dependent issues and systematics, such as the type, distribution and
amount of dark matter et cetera.

This renders potentials dominated by $V \sim m^2 \phi^2$
unsuitable for chameleonic dark energy model building, as follows.
For a quadratic potential, the environmental minimum generated by
couplings to nonrelativistic matter lies at $\frac{\alpha
\phi_*}{M_4} \, e^{-\alpha \phi_*/M_4} \sim \frac{\alpha^2
\rho}{M_4^2 m^2}$, and the scalar mass is $m^2_\phi = m^2 +
\frac{\alpha^2 \rho}{M_4^2} e^{\alpha \phi_*/M_4}$. If this field
were quintessence, at cosmological scales where $\rho \sim M^2_4
H_0^2$ its mass $m^2_\phi = m^2 + \alpha^2 H^2_0 e^{\alpha
\phi_*/M_4}$ must be smaller than $H_0^2$, as explained above,
implying the same for the vacuum mass, $m^2 < H_0^2$. But then,
the environmental minimum would be at $\frac{\alpha \phi_*}{M_4}
\, e^{-\alpha \phi_*/M_4} > \frac{\alpha^2 \rho}{M^2_4 H^2_0} \gg
1$ for relevant environments, implying that $\alpha \phi_*/M_4$
and therefore $G_{N \, eff}$ change too much with variations of
$\rho$. Conversely, one could suppress variations of the effective
Newton's constant either by taking $m^2 \gg H_0^2$, or by taking
$\alpha \rightarrow 0$, but then either the field $\phi$ should be
integrated out at scales below $m$ and cannot be quintessence, or
it would altogether decouple from matter and cease to behave as a
chameleon.

Hence other potentials must be considered. Various specific examples
were discussed in \cite{khuwe}-\cite{fuz}. The works \cite{khuwe}-\cite{brawe} 
employed vacuum potentials
that can be approximated as powers $V \sim \frac{\lambda}{n}
\phi^n$ for $n \ne 2$ (positive or negative!), yielding
$V_{eff}(\phi) = \frac{\lambda}{n} \phi^n + \frac12 \rho e^{\alpha
\phi/M_4}$. After adjusting the coupling $\lambda$ to satisfy
$\alpha \phi_* \ll M_4$ at the minimum, that prevents large
variations of $G_{N \, eff}$, the effective minimum is at $\phi_*
\simeq (\frac{\alpha}{2M_4 \lambda})^{1/(n-1)} \rho^{1/(n-1)}$.
Around it, the scalar mass is dominated by $\partial_\phi^2 V$ at
the minimum, for $n \ge 2$, and is
\be m^2_\phi \simeq (n-1) \lambda^{1/(n-1)} \Bigl(
\frac{\alpha}{M_4} \Bigr)^{\frac{n-2}{n-1}} \,
\rho^{\frac{n-2}{n-1}} \, . \label{envmass} \ee
This formula breaks down for linear potentials with $n=1$, where
the correct derivation yields $\gamma = 1/2$. Indeed, for the
linear potential $V = V_0 - q \phi$, $\partial^2 V = 0$ and so the
scalar mass is entirely an environmental effect: $m^2_\phi \simeq
\frac{\alpha^2 \rho}{M_4^2} e^{\alpha \phi/M_4} \sim \rho$. Thus
generically
\be
m_\phi \propto \rho^{\gamma} \, ,
\label{massscaling}
\ee
where $\gamma = \frac{n-2}{2(n-1)}$, or $\gamma = 1/2$ for $n=1$.
When the matter couplings of $\phi$ are of the gravitational
strength, $\alpha \sim 1$, this means that for all reasonable
power law potentials, with integer powers, once the environmental
mass $m_\phi$ is fixed by the laboratory bounds on Earth, $m_\phi
\ga 10^{-3} \, {\rm eV}$, for $\rho_{Earth} \sim {\rm g}/{\rm
cm}^3 \sim 10^{21} \, {\rm eV}^4$, it can decrease at most by a
factor of
\be \Bigl( \frac{M^2_4 H_0^2}{\rho_{Earth}} \Bigr)^{\gamma} \simeq
10^{-33 \gamma} \, , \label{rescaling} \ee
as the energy density changes to the cosmological background
density. Having started at $m_\phi \ga 10^{-3} \, {\rm eV}$, the
effective environmental mass can therefore decrease down only to
$m_\phi \ga 10^{-3 - 33 \gamma} \, {\rm eV}$. For $\phi$ to be
quintessence, suspended in slow roll on a potential slope at very
large distance scales, this must be smaller than $H_0 \sim
10^{-33} \, {\rm eV}$, which therefore requires $\gamma \ge 1$.
Otherwise, the field $\phi$ will be too heavy to have any
significant dynamics at the horizon scale, and dark energy must
come from other quarters, if at all\footnote{One can check that
similar arguments also apply to, for example, exponential
potentials. In that case, one also finds that the quintessence
mass scales as (\ref{massscaling}) with $\gamma < 1$ and that
generically it is impossible to keep $|\alpha \phi_*| < M_4$ over
a wide range of density variations.}, which is what happens with all 
integer powers.

The exception to this conclusion is the logarithmic potential $V
\sim \ln \phi$. To see that it evades the arguments above, we note
that to get the mass of $\phi$ for this case, we can take the
limit of Eq. (\ref{envmass}) when $n \rightarrow 0^-$. Then,
$\gamma \rightarrow 1$, and so $m_\phi \sim  \rho$. In this case,
the effective mass will change by the full range of density ratio
between the cosmological and terrestrial scales, spanning over
thirty orders of magnitude. Hence, the logarithmic potential {\it
can} give us a chameleonic, or changeling quintessence which could
have evaded the laboratory searches for deviations from Newton's
law at the current level of sensitivity, but may remain close to
the bounds, within the reach of the future tests.

\begin{figure}[thb]
\centerline{\includegraphics[width=9.5cm, height=7cm,
angle=0]{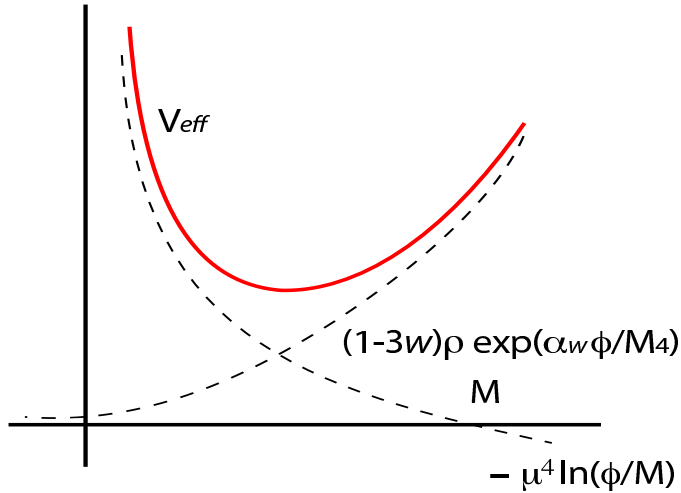}} \caption{{Environmental effective
potential.} \label{fig1}}
\end{figure}
To explore the physics of our logarithmic dark energy changeling,
we now turn to a specific model. Suppose that the vacuum potential
of the scalar is $V = - \mu^4 \ln\Bigl(\phi/M\Bigr)$,
where\footnote{We will work with $M$ not much greater than the
Planck scale, to comply with the arguments about absence of
ultraweak forces and trans-Planckian cutoffs in plausible UV
completions of gravity \cite{tomnima}. } $M \ga M_4$. Potentials
like this may arise in theories with two conical extra dimensions,
after their stabilization \cite{conepots}, or by integrating out
some heavy fields which couple to the scalar $\phi$, such as in
the {\it MaVaNs} models of \cite{mavans} which also employ
logarithmic potentials. However in contrast to {\it MaVaNs}, our
$\phi$ couples universally to all `light' matter, and dwells in a
different regime, as exemplified by the sign and our choice of the
scale $M$. As we will see later, we will need $\mu \sim 10^{-3} \,
{\rm eV}$, which is usual for quintessence models that can fit the
data. We won't commit to any particular mechanism explaining how
such scales may arise (and in particular why there aren't larger
corrections to $V$, which of course is the full cosmological
constant problem that we can't solve yet \cite{wein}), instead
focusing on their implications for observations. Nevertheless, we note that
obtaining such potentials may only require tunings in the gravitational sector,
if the scalar $\phi$ is a Brans-Dicke-like field, obeying weak equivalence principle,
since there exists a Brans-Dicke frame to which matter couples universally.
Then, the
effective potential including the environmental correction from a
medium obeying equation of state $p/\rho = w$ is
\be V_{eff}(\phi) = - \mu^4 \ln\Bigl(\frac{\phi}{M}\Bigr) + (1-3w)
\rho \, e^{\alpha_w \phi/M_4} \, . \label{logpot} \ee
It is given in Figure (\ref{fig1}) for a fixed value of $\rho$,
and for $w<-1/3$. For $w=-1/3$ the environmental term is absent,
whereas for $w>1/3$ it changes sign and convexity (since $\alpha_w
= (1-3w) \alpha$).

Let's examine cosmological history of such a theory, and see what
are its predictions. We will work with the assumption that our
universe was shaped by inflation, at some high scale $\Lambda \gg
\mu^4$ and with $w \simeq -1$. Then, during inflation, the scalar
field $\phi$ is controlled by $V_{eff}(\phi) = - \mu^4
\ln\Bigl(\frac{\phi}{M}\Bigr) + 4 \Lambda \, e^{4\alpha \phi/M_4}
$. The minimum of this potential is at $\frac{\alpha \phi_*}{M_4}
\simeq \frac{\mu^4}{16 \Lambda} \ll 1$, and the effective scalar
mass there is $m_\phi^2 \simeq \frac{256 \alpha^2 \Lambda^2}{M_4^2
\mu^4} \gg H^2_{inflation}$. In fact this is generically so large
that the field $\phi$ is completely non-dynamical during
inflation. It is frozen out extremely efficiently.

At the end of inflation, the energy in the inflaton potential
$\Lambda$ will be converted into radiation. During this stage, the
universe will rapidly become radiation-dominated, with
$\rho_{radiation} \gg \rho_{matter}$. Ignoring the possibility of
massive matter decays, we can place a bound on the ratio of
$\rho_{radiation}/\rho_{matter}$ by scaling it up from
matter-raditation equality to the reheating temperature:
$\rho_{radiation}/\rho_{matter} \ga T_{reheating}/{\rm eV}$, which
can be as high as $10^{20}$ or so. In reality this ratio will be
even higher because many of the nonrelativistic species today will
have behaved as relativistic particles in the early universe. Now,
the presence of massive particle species may generate a different
effective potential for $\phi$, shifting the location of the
environmental minimum. The environmental potential coming from
nonrelativistic species is $V_{eff}(\phi) = - \mu^4
\ln\Bigl(\frac{\phi}{M}\Bigr) + \rho_{matter} \, e^{\alpha
\phi/M_4} $, with a minimum at $\frac{\alpha \phi_*}{M_4} \simeq
\frac{\mu^4}{\rho_{matter}}$, and a mass around it $m^2_\phi
\simeq \frac{\alpha^2 \rho_{matter}^2}{M_4^2 \mu^4} \simeq
\frac{\alpha^2
\rho_{matter}}{\mu^4}\frac{\rho_{matter}}{\rho_{radiation}}
H^2_{radiation}$. During the radiation phase $\rho_{radiation} \gg
\rho_{matter}$, and so $m^2_{\phi} \ll H^2_{radiation}$. This
minimum, if at all present, will be too shallow to affect
cosmological dynamics of $\phi$.

Thus we can ignore $\rho_{matter}$ during the radiation epoch. The
effective potential for $\phi$ changes to the pure logarithmic
term, where the field is massless and initially close to the
origin, where inflation left it: $\phi_{inflation} \simeq
\frac{\mu^4 M_4}{16 \Lambda \alpha}$. However generically the
field will have a lot of kinetic energy after being released from
its inflationary state. To see that, introduce $\tilde \rho =
\Lambda e^{4 \alpha \phi/M_4}$ as the total energy density during
inflation. Just before the end of inflation, where $\Lambda$
starts to decay, the time variation of $\Lambda$ will pull along
$\phi$, $ \dot {\tilde \rho} \ga \frac{4 \alpha \dot \phi}{M_4}
\tilde \rho$, whence $\dot \phi \la \frac{\dot
{\tilde\rho}}{4\alpha \tilde \rho} M_4$. With efficient reheating
we can estimate $\frac{\dot {\tilde \rho}}{\tilde \rho} \simeq
H_{inflation}$, so that $\dot \phi \la \frac{M_4
H_{inflation}}{4\alpha}$, or $\dot \phi^2 \la \frac{\Lambda}{48
\alpha^2}$. While by no means precise, this argument at least shows that at
the end of inflation, the field $\phi$ will generically convert a
significant fraction of vacuum energy into its kinetic energy, by
the universality of its couplings to all types of matter and
equipartition of energy. The precise amount would depend on the
model of inflation and reheating. Having so much kinetic energy
after inflation is not dangerous for cosmology because it will
dissipate quickly due to Hubble friction/redshift. Since we can
neglect nonrelativistic matter at this stage, and because the
potential energy density at this time is $V \sim \mu^4 \ll
\Lambda$, we can in fact ignore the effective potential
altogether. As a result the field will evolve as a pure massless
mode in a radiation-dominated universe, where it will stop more or
less after a Hubble time, travelling a distance $\Delta \phi \sim
\dot \phi_{initial}/H_{inflation} \la \frac{M_4}{4\alpha} \gg
\phi_{initial}$ before it stops \cite{freezing}. At that point, it
will have an expectation value $\phi \la  \frac{M_4}{4\alpha}$, a
tiny potential energy, $V \la \mu^4 \ln\bigl(\frac{4\alpha M}{M_4}
\bigr)$, and a tiny mass\footnote{For as long as $\rho_{matter} >
\mu^4$.} $m^2_\phi \simeq \frac{\alpha^2 \rho_{matter}}{M_4^2} \ll
H^2_{radiation}$, giving a slightly stronger effective
gravitational coupling $G_{N \, eff}$ to matter than to radiation,
by at most a factor of about $\la e^{1/4} \sim 1.28$ or so. For
the rest of the radiation era, the field will simply just wait
there.

We should comment here on the implications of the enhancement of
$G_{N \, eff}$ for Big Bang nucleosynthesis. A difference between
the value of Newton's constant in the early universe, and
specifically at the time of BBN and its value measured presently
in terrestrial experiments would affect relic abundances, and so
BBN gives us strong limits on the variation of $G_N$
\cite{garyold}. However, our calculated maximal value of $G_{N \,
eff}$ above, is the value of Newton's constant at nucleosynthesis
as seen by nonrelativistic particles, with masses $m \gg {\rm
MeV}$ at that time. Indeed, we recall that -- as illustrated in
e.g. Eq. (\ref{effpot}) -- the effective Newton's constant which a
species sees is $G_{N \, eff} \sim \frac{1}{M_4^2} e^{\alpha_w
\phi/M_4}$, where $\alpha_w = (1-3w) \alpha$. Thus the
relativistic particles, which are controlling the expansion rate
of the universe at that time, would feel an effective Newton's
constant much closer to its terrestrial value. Even the maximal
value which we estimated above, felt by heavy particles, may be
consistent with the new BBN bounds on $\Delta G_N/G_{N \, 0}$ that
allow it to be $\sim 20 \%$  \cite{krauss,keith}, although
stronger bounds may be inferred from different data \cite{garys}.
Hence BBN data may probe this
aspect of our model, similarly to what happens in 
general scalar-tensor theories \cite{dano}. 
This should be explored in more detail. We
need to also stress that the bounds from Oklo are easy to comply
with. By the time the Oklo reactor started, the field would have settled
into its terrestrial minimum, pulling $G_{N \, eff}$ down to its
familiar value.

After radiation-matter transition, $\propto \rho_{matter}$ term in the
effective potential will be of the order of $M_4^2 H^2$. The
environmental minimum for $\phi$ at the largest scales will become
more prominent, and its location, as previously calculated, will
be at $\frac{\alpha \phi_*}{M_4} \simeq
\frac{\mu^4}{\rho_{matter}}$. Now, in dilute universe before
structure formation, but after radiation-matter transition,
$\rho_{matter}$ will be below ${\rm eV}^4$, approaching $\mu^4$
from above. This means, that the minimum has been shifting towards
the Planckian values, where the field has been laying in wait.
Yet, as long as $\alpha < 1$, the field will not shift from where
it went during the radiation epoch. The reason is that as long as
$\rho_{matter} > \mu^4$, it's mass is still given by $m^2_\phi
\simeq \frac{\alpha^2\rho_{matter}}{M^2_4}$. So by arranging
$\alpha < 1/\sqrt{3}$, we can still keep $m^2_\phi <
\frac{\rho_{matter}}{3M_4^2} = H^2_{matter}$, holding the field up
on the logarithmic slope by Hubble friction.

On the other hand, at shorter scales structure will begin to form
around the primordial gravitational wells generated during inflation, where
matter will agglomerate and the local matter density will increase
manyfold over the uniform background value. In these regions, the
environmental minima for $\phi$ will be closer to the origin and
deeper, with $m_\phi^2 \gg H^2_{matter}$. Hence 
where collapse began the field $\phi$ will fall back to the
environmental minimum, oscillating around it instead of sticking to its
post-inflationary value. In these regions, therefore, the field will behave as a component 
of cold dark matter, and its uniform energy density inside the region 
will begin to redshift as $\sim 1/a^3$, yielding
the scaling of $\phi \sim 1/a^{3/2}$, similarly
to unified dark matter models \cite{udm,fuz}. This stage of evolution
can reduce the field value by as much as $\sim 10^7$ inside large scale overdensities. 
Moreover, at shorter scales gravitational cooling of the field \cite{gravcool} will lead 
to the collapse of the field energy to the core of the 
distribution, as in scalar field dark matter models, and to virialization 
with collapsing matter \cite{matos}. 
This will further reduce the value of the scalar field around the central overdensity 
to $\phi \ll M_4$, sweeping it into the center. Finally, where the matter 
overdensity reaches the scales of $\rho \sim 10^{6} \, {\rm eV}^4$ and beyond,  
the field mass will be $m_\phi \ga 10^{-16} \, {\rm eV}$, so that the leftover field 
oscillations in time will occur at frequencies $\ga {\rm sec}^{-1}$ about the minimum, 
so that we may replace it with its time average, $\phi_*$. 
So the long range effects of fields in these regions should be suppressed 
by the conspiracy between its environmental mass and
the thin shell effect. Overall however one must be careful about
picking the field boundary conditions in determining the long range forces
as these depend sensitively on the evolution of matter and field distributions.
To set up the long range fields, in general 
one therefore needs to look at the full history of the system. 
It is also possible that the field may leave some imprint in the
large scale structure, since it will be more active in the
beginning of the collapse. The precise description of these
imprints is beyond the scope of this work, but we expect that
because the imprints arise due to stronger gravity, they may
affect our determination of dark matter abundance, leading us to
overestimate the abundance of dark matter in structures which are
at an early stage of their formation. Presumably this may lead to the possibility of
direct astronomical tests and it would be interesting to develop
further. 

Back at cosmological scales, the evolution will eventually dilute
$\rho_{matter}$ to below $\mu^4$. At this time, the universe will
become dominated by the small residual potential energy in the
field, $V \la \mu^4 \ln(\frac{4\alpha M}{M_4}) \sim \mu^4$. The
environmental minimum will shift to $\frac{\alpha \phi_*}{M_4} >
1$. However, the effective field mass at the largest scales will
change to $m_\phi^2 \simeq \frac{\mu^4}{\phi^2}$, which is
initially $m^2_\phi \simeq 16 \frac{\alpha^2 \mu^4}{M_4^2}$. So
the field will remain away from the minimum, and will start to
slowly roll towards it as $\rho_{matter}$ dips below $\mu^4$. To
ensure that the universe accelerates right away, we need to
enforce Eq. (\ref{mdark}). At this time, Eq. (\ref{mdark})
translates to
\be
\alpha \la \frac{1}{4\sqrt{3}} \, .
\label{alpha}
\ee
Similarly, we must also demand that $V>0$, which implies $4\alpha
M > M_4$, and that the period of acceleration lasts at least an
e-fold or so, $\Delta t \ga 1/H_0$. A stronger bound on $M$ comes
about as follows. As time goes on and $\phi$ rolls down the
logarithmic slope, the slow roll will improve, as $m^2_\phi$ is
{\it decreasing} with the increase of $\phi$, as $m^2_\phi \simeq
\frac{\mu^4}{\phi^2}$. Thus solving the field equations
(\ref{hubble})-(\ref{fieldeq}) in the slow roll regime, we find
that
\be \frac{\mu^2 M_4}{\sqrt{3}} \Delta t \simeq
\int^{\phi}_{\phi_0} d\phi \phi \,
\ln^{1/2}\bigl(\frac{M}{\phi}\bigr) \, , \label{phievol} \ee
where $\phi_0$ is the value of $\phi$ at the beginning of
acceleration, $\phi_0 \la \frac{M_4}{4\alpha}$. The integral is
extremized by taking $\phi_0 = \frac{M_4}{4\alpha}$, and $\phi =
M$, because the log potential will vanish there, and so if there
are no higher order corrections that can prevent the potential
from going negative, acceleration will only last until $\phi$
reaches $M$. Beyond that, acceleration will cease, and in fact the
universe may even collapse, as has been recently studied in
\cite{andreiren}. So substituting $\phi = M e^{-x/2}$, the
integral reduces to $\frac{M^2}{2^{3/2}} \int^{2\ln(\frac{4\alpha
M}{M_4})}_0  dx \sqrt{x} \, e^{-x}$. By using $4\alpha M > M_4$
and Eq. (\ref{alpha}), we can maximize it with an Euler gamma
function $\Gamma(\frac32)$. The error is tolerable, as one can
verify by using the saddle point approximation. Thus, the total
duration of the late accelerating phase cannot be longer than
$\Delta t \simeq \sqrt{\frac{3\pi}{32}} \frac{M^2}{\mu^2 M_4}$.
The logarithmic plateau needs to be wide enough to accommodate at
least an e-fold of inflation during this time, which, after
setting $H_0 \simeq \frac{\mu^2}{\sqrt{3} M_4}$ and requiring $H_0
\Delta t \ga 1$, implies that
\be
M \ga  \bigl( \frac{32}{\pi} \bigr)^{1/4} \, M_4 \simeq 1.78 \, M_4 \, .
\label{mass}
\ee
This will suffice to explain the observed cosmic acceleration. We
note that the criticisms of the {\it MaVaNs} model \cite{zalda} (see also \cite{reex})
are easily circumvented here, since $\phi$ is in the slow roll
regime, independently of the matter terms from the onset of
acceleration.

Now if we don't take $M$ too large, avoiding UV cutoffs much
higher than the Planck scale \cite{tomnima}, the scalar may have
matter couplings to within an order of magnitude of the
gravitational couplings. To see it we can combine (\ref{alpha})
and inequality $4\alpha M > M_4$ into
\be
\frac{M_4}{4M} < \alpha \la \frac{1}{4\sqrt{3}} \, .
\label{coupin}
\ee
Since the scalar coupling to matter is governed by 
\be
g_{\phi} \sim \frac{\alpha}{M_4} \, ,
\label{gphi}
\ee
and its mass in terrestrial environments, where $\rho_{matter} \gg \mu^4$, is
\be 
m_\phi \sim \frac{\alpha \rho_{matter}}{M_4 \mu^2} \sim \frac{\alpha}{10} \, {\rm eV} \, ,
\label{massphiter}
\ee
when $M$ is not too large there remains a chance that $\phi$ could 
be within the reach of the future
laboratory searches, after further improvements in sensitivity.
Moreover, in this case $\phi$ will be rolling noticeably after an
e-fold or so. Hence it would behave as $w\ne -1$ dark energy.

To summarize, we have delineated a dark energy model which, while
tuned as it stands now, can be tested at several different
observational fronts. It is based on a light scalar, with slightly
sub-gravitational couplings to matter and a mass which depends on
the environmental energy density. Outside of dense matter
distributions this field will be light, and may yield significant
long range effects. In particular, if it has logarithmic
self-interaction potential, like those that can arise in theories
with conical extra dimensions \cite{conepots,sliver}, or is
generated radiatively \cite{mavans}, it can be quintessence, with
mass $m_\phi \la H_0$. At the largest scales however, this field
will couple to matter in contrast to typical quintessence models,
albeit slightly more weakly than gravity. In the early universe it
will have an expectation value that is larger than in the
terrestrial minima, which would make gravity slightly stronger.
This can have consequences for BBN. During structure formation,
before the field decouples in deeper potential wells around denser
matter distributions, it may have affected cosmic structures. We
have not analyzed this in detail here, and it would be very
interesting to determine precisely what kind of signatures can
arise. They may imitate an excess in the amount of dark matter.
Finally, at the scales governing terrestrial physics, this field
will become sufficiently massive so that its long range force may be
suppressed by the thin shell effect discussed in the context of
chameleons.
Hence it may have avoided detection to date. However,
its effects may be probed by future searches for sub-millimeter
corrections to gravity. We believe that this represents an
interesting framework for testing gravity and dark energy in a
correlated manner. Testing models which involve correlations
between modifications of gravity at short and long scales will
probe the robustness of General Relativity and its greatest
failure, the cosmological constant. It is therefore important to
scrutinize such ideas further. Perhaps, ultimately, we might even
end up getting surprised!

\vskip2.5cm

{\bf \noindent Acknowledgements}

\smallskip

We thank A. Albrecht, L. Knox, K. Olive, L. Sorbo, G. Steigman, S. Watson and
especially  J. Khoury and J. A. Tyson for interesting conversations. This work
was supported in part by the DOE Grant DE-FG03-91ER40674 and in
part by a Research Innovation Award from the Research Corporation.



\begin{thebibliography}{99}

\bibitem{wein}
S.~Weinberg,
Rev.\ Mod.\ Phys.\  {\bf 61}, 1 (1989).

\bibitem{whynow}
N.~A.~Bahcall, J.~P.~Ostriker, S.~Perlmutter and P.~J.~Steinhardt,
Science {\bf 284}, 1481 (1999);
%
I.~Zlatev, L.~M.~Wang and P.~J.~Steinhardt,
Phys.\ Rev.\ Lett.\  {\bf 82}, 896 (1999);
%
N.~Arkani-Hamed, L.~J.~Hall, C.~F.~Kolda and H.~Murayama,
Phys.\ Rev.\ Lett.\  {\bf 85}, 4434 (2000).

\bibitem{beane}
S.~R.~Beane,
Gen.\ Rel.\ Grav.\  {\bf 29}, 945 (1997).

\bibitem{dapive}
T.~Damour, F.~Piazza and G.~Veneziano,
Phys.\ Rev.\  D {\bf 66}, 046007 (2002);
Phys.\ Rev.\ Lett.\  {\bf 89}, 081601 (2002).

\bibitem{mavans}
R.~Fardon, A.~E.~Nelson and N.~Weiner,
JCAP {\bf 0410}, 005 (2004);
JHEP {\bf 0603}, 042 (2006);
D.~B.~Kaplan, A.~E.~Nelson and N.~Weiner,
Phys.\ Rev.\ Lett.\  {\bf 93}, 091801 (2004).
 
\bibitem{gwz}
P.~Gu, X.~Wang and X.~Zhang,
Phys.\ Rev.\  D {\bf 68}, 087301 (2003);
X.~J.~Bi, B.~Feng, H.~Li and X.~m.~Zhang,
Phys.\ Rev.\  D {\bf 72}, 123523 (2005).

\bibitem{lawrence}
L.~J.~Hall, Y.~Nomura and S.~J.~Oliver,
Phys.\ Rev.\ Lett.\  {\bf 95}, 141302 (2005);
R.~Barbieri, L.~J.~Hall, S.~J.~Oliver and A.~Strumia,
Phys.\ Lett.\  B {\bf 625}, 189 (2005).

\bibitem{quint}
S.~M.~Carroll,
Phys.\ Rev.\ Lett.\  {\bf 81}, 3067 (1998);
C.~F.~Kolda and D.~H.~Lyth,
Phys.\ Lett.\ B {\bf 458}, 197 (1999);
T.~Chiba,
Phys.\ Rev.\ D {\bf 60}, 083508 (1999).

\bibitem{laboratory}
E.~G.~Adelberger, B.~R.~Heckel, S.~Hoedl, C.~D.~Hoyle, D.~J.~Kapner and A.~Upadhye,
Phys.\ Rev.\ Lett.\  {\bf 98}, 131104 (2007);
D.~J.~Kapner, T.~S.~Cook, E.~G.~Adelberger, J.~H.~Gundlach, B.~R.~Heckel, C.~D.~Hoyle and H.~E.~Swanson,
Phys.\ Rev.\ Lett.\  {\bf 98}, 021101 (2007);
see also E.~G.~Adelberger, B.~R.~Heckel and A.~E.~Nelson,
Ann.\ Rev.\ Nucl.\ Part.\ Sci.\  {\bf 53}, 77 (2003).

\bibitem{dano}
T.~Damour and K.~Nordtvedt,
Phys.\ Rev.\ Lett.\  {\bf 70}, 2217 (1993);
Phys.\ Rev.\  D {\bf 48}, 3436 (1993).

\bibitem{dapol}
T.~Damour and A.~M.~Polyakov,
Nucl.\ Phys.\  B {\bf 423}, 532 (1994);
Gen.\ Rel.\ Grav.\  {\bf 26}, 1171 (1994).

\bibitem{khuwe}
J.~Khoury and A.~Weltman,
Phys.\ Rev.\ Lett.\  {\bf 93}, 171104 (2004);
Phys.\ Rev.\  D {\bf 69}, 044026 (2004).

\bibitem{dgs}
G.~R.~Dvali, G.~Gabadadze and M.~A.~Shifman,
Mod.\ Phys.\ Lett.\  A {\bf 16}, 513 (2001).

\bibitem{gukhu}
S.~S.~Gubser and J.~Khoury,
Phys.\ Rev.\  D {\bf 70}, 104001 (2004).

\bibitem{brawe}
P.~Brax, C.~van de Bruck, A.~C.~Davis, J.~Khoury and A.~Weltman,
Phys.\ Rev.\  D {\bf 70}, 123518 (2004).

\bibitem{weicai}
H.~Wei and R.~G.~Cai,
Phys.\ Rev.\  D {\bf 71}, 043504 (2005);
T.~Biswas, R.~Brandenberger, A.~Mazumdar and T.~Multamaki,
Phys.\ Rev.\  D {\bf 74}, 063501 (2006).

\bibitem{udm}
A.~Y.~Kamenshchik, U.~Moschella and V.~Pasquier,
Phys.\ Lett.\  B {\bf 511}, 265 (2001).

\bibitem{fuz}
A.~Fuzfa and J.~M.~Alimi,
Phys.\ Rev.\ Lett.\  {\bf 97}, 061301 (2006);
Phys.\ Rev.\  D {\bf 73}, 023520 (2006);
arXiv:astro-ph/0702478.

\bibitem{tomnima}
M.~Dine,
arXiv:hep-th/0107259;
T.~Banks, M.~Dine, P.~J.~Fox and E.~Gorbatov,
JCAP {\bf 0306}, 001 (2003);
N.~Arkani-Hamed, L.~Motl, A.~Nicolis and C.~Vafa,
arXiv:hep-th/0601001.

\bibitem{conepots}
N.~Arkani-Hamed, L.~J.~Hall, D.~R.~Smith and N.~Weiner,
Phys.\ Rev.\  D {\bf 62}, 105002 (2000);
M.~J.~May and R.~Sundrum,
Phys.\ Rev.\  D {\bf 69}, 104010 (2004).

\bibitem{freezing}
N.~Kaloper and K.~A.~Olive,
Astropart.\ Phys.\  {\bf 1}, 185 (1993);
A.~A.~Tseytlin and C.~Vafa,
Nucl.\ Phys.\  B {\bf 372}, 443 (1992);
T.~Barreiro, B.~de Carlos and E.~J.~Copeland,
Phys.\ Rev.\  D {\bf 58}, 083513 (1998);
L.~Kofman, A.~Linde, X.~Liu, A.~Maloney, L.~McAllister and
E.~Silverstein,
JHEP {\bf 0405}, 030 (2004);
S.~Watson,
Phys.\ Rev.\ D {\bf 70}, 066005 (2004);
R.~Brustein, S.~P.~de Alwis and P.~Martens,
Phys.\ Rev.\  D {\bf 70}, 126012 (2004);
N.~Kaloper, J.~Rahmfeld and L.~Sorbo,
Phys.\ Lett.\  B {\bf 606}, 234 (2005);
T.~Barreiro, B.~de Carlos, E.~Copeland and N.~J.~Nunes,
Phys.\ Rev.\  D {\bf 72}, 106004 (2005);
S.~Cremonini and S.~Watson,
Phys.\ Rev.\ D {\bf 73}, 086007 (2006);
B.~Greene, S.~Judes, J.~Levin, S.~Watson and A.~Weltman,
arXiv:hep-th/0702220.

\bibitem{garyold}
G.~Steigman,  Nature {\bf 261}, 479 (1976);
A.~M.~Boesgaard and G.~Steigman,
Ann.\ Rev.\ Astron.\ Astrophys.\  {\bf 23}, 319 (1985).

\bibitem{krauss}
C.~J.~Copi, A.~N.~Davis and L.~M.~Krauss,
Phys.\ Rev.\ Lett.\  {\bf 92}, 171301 (2004).

\bibitem{keith}
R.~H.~Cyburt, B.~D.~Fields, K.~A.~Olive and E.~Skillman,
Astropart.\ Phys.\  {\bf 23}, 313 (2005).

\bibitem{garys}
G.~Steigman,
 Int.\ J.\ Mod.\ Phys.\  E {\bf 15}, 1 (2006).

\bibitem{gravcool}
E.~Seidel and W.~M.~Suen,
Phys.\ Rev.\ Lett.\  {\bf 66}, 1659 (1991);
Phys.\ Rev.\ Lett.\  {\bf 72}, 2516 (1994).

\bibitem{matos}
T.~Matos and L.~A.~Urena-Lopez,
Class.\ Quant.\ Grav.\  {\bf 17}, L75 (2000);
T.~Matos and L.~A.~Urena-Lopez,
Phys.\ Rev.\  D {\bf 63}, 063506 (2001);
M.~Alcubierre, F.~S.~Guzman, T.~Matos, D.~Nunez, L.~A.~Urena-Lopez and P.~Wiederhold,
Class.\ Quant.\ Grav.\  {\bf 19}, 5017 (2002);
F.~S.~Guzman and L.~A.~Urena-Lopez,
Phys.\ Rev.\  D {\bf 69}, 124033 (2004).


\bibitem{andreiren}
R.~Kallosh, J.~Kratochvil, A.~Linde, E.~V.~Linder and M.~Shmakova,
JCAP {\bf 0310}, 015 (2003).

\bibitem{zalda}
N.~Afshordi, M.~Zaldarriaga and K.~Kohri,
Phys.\ Rev.\  D {\bf 72}, 065024 (2005).

\bibitem{reex}
O.~E.~Bjaelde, A.~W.~Brookfield, C.~van de Bruck, S.~Hannestad, D.~F.~Mota, L.~Schrempp and D.~Tocchini-Valentini,
arXiv:0705.2018 [astro-ph].

\bibitem{sliver}
N.~Kaloper,
arXiv:hep-th/0702206.

\end{thebibliography}
\end{document}